\documentclass[aps,prl,twocolumn,groupedaddress,showpacs,amsmath,amssymb]{revtex4}

\usepackage{graphicx}
\usepackage{dcolumn}
\usepackage{bm}

\begin{document}

\title{Coincidence of magnetic and valence quantum critical points in CeRhIn$_{5}$ under pressure}

\date{\today}
\author{Z. Ren$^{1}$$^{,}$$^{2}$}
\email{zhi.ren@wias.org.cn}
\author{G. W. Scheerer$^{1}$}
\author{D. Aoki$^{3}$$^{,}$$^{4}$}
\author{K. Miyake$^{5}$}
\author{S. Watanabe$^{6}$}
\author{D. Jaccard$^{1}$}
\affiliation{$^{1}$DQMP - University of Geneva, 24 Quai Ernest-Ansermet, 1211 Geneva 4, Switzerland}
\affiliation{$^{2}$Institute for Natural Sciences, Westlake Institute for Advanced Study, 18 Shilongshan Road, Hangzhou, P. R. China}
\affiliation{$^{3}$Univ. Grenoble Alpes, Commissariat $\grave{a}$ l$^{\prime}$Energie Atomique et aux Energies Alternatives (CEA), Institut Nanosciences et Cryogenie (INAC)-PHELIOS, F-38000 Grenoble, France}
\affiliation{$^{4}$Institute for Materials Research, Tohuko University, Oarai, Ibaraki 311-1313 Japan}
\affiliation{$^{5}$Center for Advanced High Magnetic Field Science, Osaka University, Toyonaka 560-0043, Japan}
\affiliation{$^{6}$Department of Basic Sciences, Kyushu Institute of Technology, Kitakyushu 804-8550, Japan}

\begin{abstract}
We present accurate electrical resistivity measurements along the two principle crystallographic axes of the pressure-induced heavy-fermion superconductor CeRhIn$_{5}$ up to 5.63 GPa. For both directions, a valence crossover line is identified in the $p$$-$$T$ plane and the extrapolation of this line to zero temperature coincides with the collapse of the magnetic ordering temperature. Furthermore, it is found that the $p$$-$$T$ phase diagram of CeRhIn$_{5}$ in the valence crossover region is very similar to that of CeCu$_{2}$Si$_{2}$. These results point to the essential role of Ce-4$f$ electron delocalization in both destroying magnetic order and realizing superconductivity in CeRhIn$_{5}$ under pressure.
\end{abstract}

\maketitle
\maketitle

\section{I. Introduction}

The occurrence of pressure-induced superconductivity (SC) in Ce-based heavy-fermion (HF) compounds has attracted a lot of attention in the field of condensed matter physics \cite{Cebasedreview}.
For most of such materials, SC appears in the vicinity of a magnetic quantum critical point (QCP) at $p_{\rm c}$, leading to the belief that spin fluctuations are responsible for the Cooper pairing \cite{spinfluctuation}.
On the other hand, Ce-valence fluctuations may also act as the pairing glue \cite{valencefluctuation} and the corresponding critical endpoint (CEP) at $p_{\rm v}$ can be deduced by resistivity scaling analysis \cite{seyfarth}.
It is noteworthy that the relative position between $p_{\rm c}$ and $p_{\rm v}$ may vary for different systems, probably depending on the hybridization strength ($V$) between Ce-4$f$ and conduction electrons \cite{watanabe1,Miyake2}.
For example, while $p_{\rm c}$ and $p_{\rm v}$ are well separated in CeCu$_{2}$Si$_{2}$ \cite{valencefluctuation} and CeCu$_{2}$(Si$_{1-x}$Ge$_{x}$)$_{2}$ \cite{Yuan}, they are very close in CeAu$_{2}$Si$_{2}$ \cite{RenPRX}.

CeRhIn$_{5}$ belongs to the Ce-115 family, whose structure consists of alternating CeIn$_{3}$ and RhIn$_{2}$ layers stacked along the $c$-axis \cite{Ce115discovery}. At ambient pressure, the compound is a prototypical heavy-fermion antiferromagnet with a N\'{e}el temperature $T_{\rm N}$ = 3.8 K, although a signature of SC was reported at $\sim$90 mK \cite{Ce115ambientSC}.
Under pressure, the $T_{\rm N}$ of CeRhIn$_{5}$ passes through a maximum and disappears at $\sim$2 GPa, above which the antiferromagnetic order is rapidly suppressed as confirmed by the NQR measurement \cite{Ce115NQR}. Meanwhile, SC is observed over a broad pressure range with a maximum $T_{\rm c}$ of 2.3 K at $p_{\rm c}$ $\approx$ 2.5 GPa \cite{Ce115discovery}. Although experimental signatures for the existence of a QCP are found at $p_{\rm c}$, the nature of fluctuations remains under debate \cite{park,Ce115Knebel}.
In particular, de Haas-van Alphen (dHvA) measurements detect an abrupt change in the Fermi surface shape across $p_{\rm c}$ \cite{Ce115QO}, yet the resistivity above $T_{\rm c}$ remains nearly isotropic \cite{park}.
This is hardly understood within the common picture of magnetic quantum criticality.
Instead, Park \emph{et al.} attribute this $p_{\rm c}$ to a Kondo breakdown QCP, at which the whole heavy Fermi surface is destroyed \cite{park}. Alternatively, Watanabe and Miyake proposed theoretically that the coincidence of $p_{\rm c}$ and $p_{\rm v}$ is responsible for these anomalous behaviors \cite{Miyake1}.

In order to shed light on this issue, we carried out simultaneous measurements of the $a$- and $c$-axis electrical resistivity of CeRhIn$_{5}$ in a single pressure cell up to 5.63 GPa.
For both directions, analysis of the resistivity data allows us to draw the valence crossover line in the $p$$-$$T$ phase diagram.
Moreover, a scaling behavior is observed for the $a$-axis data, providing clear evidence for the existence of a CEP located at $p_{\rm v}$ = 2.6 $\pm$ 0.1 GPa and slightly negative temperature. Our results support the scenario that $p_{\rm v}$ and $p_{\rm c}$ are nearly identical in CeRhIn$_{5}$.
In addition, a comparison with CeCu$_{2}$Si$_{2}$ is made, and its implication on the pairing mechanism is discussed.

\section{II. Experimental}
High quality CeRhIn$_{5}$ and LaRhIn$_{5}$ crystals were grown by the In-flux method \cite{Ce115Knebel}.
After carefully removing the residual flux, crystals were oriented and cut along the $a$- and $c$-axis, respectively. The high-pressure experiment was performed using a Bridgman-type tungsten carbide anvil cell with Daphne oil as hydrostatic pressure medium and Pb as pressure gauge \cite{Bridgman}. Both samples and a Pb gauge were connected in series, and their resistivities were measured at temperatures from 300 down to 1.2 K by the standard four-probe method.
Throughout the experiments, the pressure gradient estimated from the width of the Pb superconducting transition was less than 0.05 GPa.
To determine better the absolute resistivity value, pressure-dependent resistivity at 292 K was extrapolated to $p$ = 0. The obtained value was corrected by the one measured at ambient condition, yielding a normalization factor for the results under pressure. Thanks to this special care, the estimated error in the absolute resistivity value is within 2\%.

\begin{figure}
\includegraphics*[width=8.5cm]{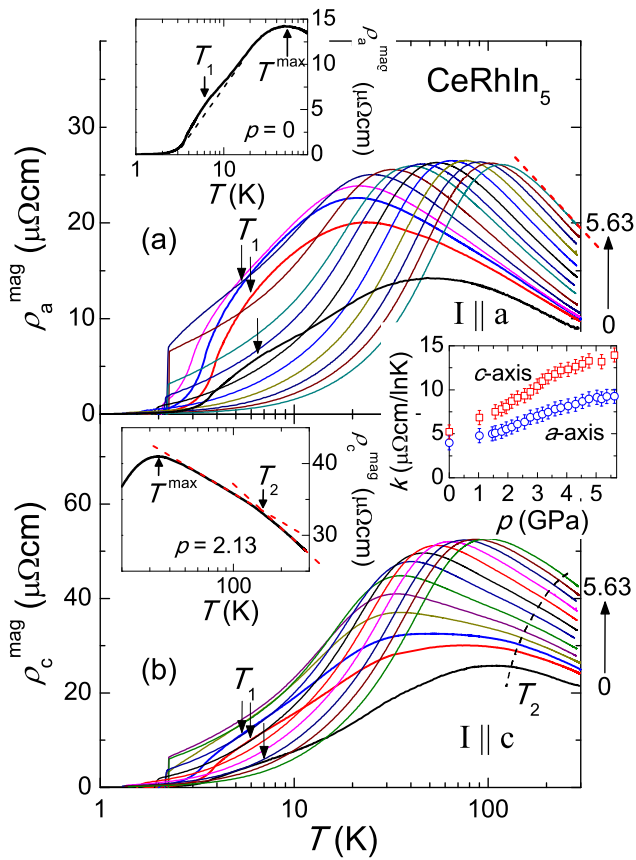}
\caption{(Color online)
Logarithmic $T$-dependence of the magnetic resistivity of CeRhIn$_{5}$ along the (a) $a$- and (b) $c$-axis under pressures up to 5.63 GPa.
The arrows and the dashed line are a guide to the eyes.
The inset of (a) shows the $a$-axis data at $p$ = 0. The resistivity maximum and bump are marked by $T^{\rm max}$ and $T_{\rm 1}$, respectively.
The dashed line is a guide to the eyes.
The inset of (b) shows the $c$-axis data at $p$ = 2.13 GPa. The two dashed lines indicate the $-$ln$T$ slope, and their intersection temperature is marked as $T_{\rm 2}$.
The inset in between (a) and (b) shows the pressure dependencies of the $-$ln$T$ slope below room temperature for both axes.
}
\label{fig1}
\end{figure}

\section{III. Results and Discussion}
Figure 1 shows the temperature dependencies of the magnetic resistivity $\rho^{\rm mag}$ = $\rho$(CeRhIn$_{5}$)$-$ $\rho$(LaRhIn$_{5}$) of CeRhIn$_{5}$ at pressures up to 5.63 GPa. The weak pressure variation of $\rho$(LaRhIn$_{5}$) is taken into account following Ref. \cite{park}. In general, the pressure evolution of $\rho^{\rm mag}$ is reminiscent of that observed in other Ce-based Kondo lattice compounds. At $p$ = 0, $\rho^{\rm mag}_{\rm a}$ exhibits a $-$ln$T$ dependence below room temperature, reflecting incoherent Kondo scattering on excited crystal field (CF) levels \cite{CFE}. Upon further cooling, a broad maximum is observed at $T_{\rm max}$ and a small bump is discernable at a lower temperature $T_{\rm 1}$, which is defined empirically as 3/4 of the temperature at which the second derivative of $\rho^{\rm mag}$ reaches a maximum. With increasing pressure, $T_{\rm 1}$ decreases modestly and becomes no longer resolvable above 1.57 GPa, while $T_{\rm max}$ first decreases then increases. In addition, a signature of magnetic ordering is observed below 1.78 GPa, while a superconducting transition occurs between 1.03 and 3.80 GPa.

As can be seen in Fig. 1(b), $\rho^{\rm mag}_{\rm c}$ behaves similarly to $\rho^{\rm mag}_{\rm a}$, except that the former displays two different $-$ln$T$ dependencies above $T_{\rm max}$. This new observation is likely due to the relatively small value of the first CF splitting energy in comparison with other Ce-based HF systems \cite{Ce115CFE}. Extrapolations of these $-$ln$T$ behaviors intersect at the temperature $T_{\rm 2}$ [inset of Fig. 1(b)], which increases with pressure. Nevertheless, the $-$ln$T$ slope $k$ near room temperature for both axes grows by nearly the same factor of 3 up to 5.63 GPa, which signifies a strong enhancement of the Kondo coupling by pressure Ref. \cite{CFE}.
\begin{figure}
\includegraphics*[width=8.5cm]{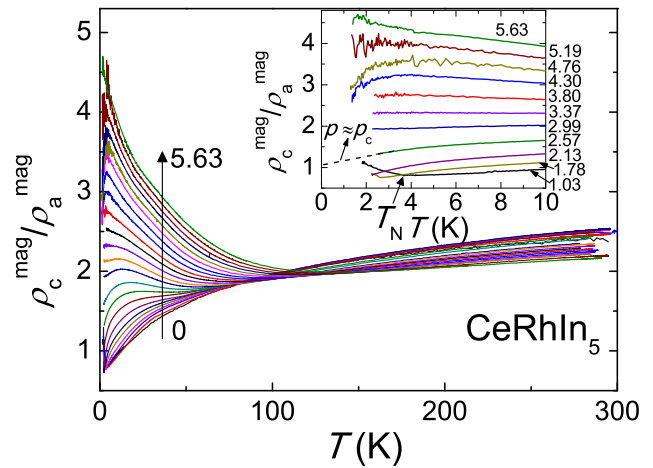}
\caption{(Color online)
Temperature dependencies of the magnetic resistivity anisotropy under pressures up to 5.63 GPa.
The inset shows a zoom of the data below 10 K at selected pressures.
The dashed line is an extrapolation of the curve at $p$ $\approx$ $p_{\rm c}$ to zero temperature.
}
\label{fig2}
\end{figure}

Figure 2 shows the anisotropy of the magnetic resistivity $\gamma_{\rm mag}$ = $\rho^{\rm mag}_{\rm c}$/$\rho^{\rm mag}_{\rm a}$ plotted as a function of temperature under pressures up to 5.63 GPa. At $p$ = 0, $\gamma_{\rm mag}$ decreases from $\sim$2.2 to $<$ 1 with decreasing temperature and shows an upturn below $T_{\rm N}$. This upturn, whose origin remain unclear at present, was not observed in the previous study \cite{C115MR}.
Under pressure, the evolution of $\gamma_{\rm mag}$ is very similar to that of \cite{park}, and exhibits qualitative difference at temperatures above and below $\sim$120 K.
For $T$ $>$ 120 K, $\gamma_{\rm mag}$ is weakly temperature and pressure dependent, and hence is likely dominated by the crystalline anisotropy.

By contrast, below $\sim$120 K, $\gamma_{\rm mag}$ increases strongly with increasing pressure and decreasing temperature.
Consequently, the temperature dependence of $\gamma_{\rm mag}$ changes its curvature from downward to upward.
At 2 K, the $\gamma_{\rm mag}$ value grows by a factor of 3 throughout the investigated pressure range  [inset of Fig.2].
This feature can likely be understood by taking into account the anisotropic hybridization between Ce-4$f$ electrons and conduction electrons \cite{anistropichybridization}. Following such an interpretation, the hybridization strength grows much faster with pressure along the $c$-axis than along the $a$-axis.
Nevertheless, at $p$ = 2.57 GPa, the $\gamma_{\rm mag}$ value extrapolated to 0 K is very close to 1, pointing to isotropic magnetic scattering around $p_{\rm c}$ \cite{park}.

\begin{figure}
\includegraphics*[width=8cm]{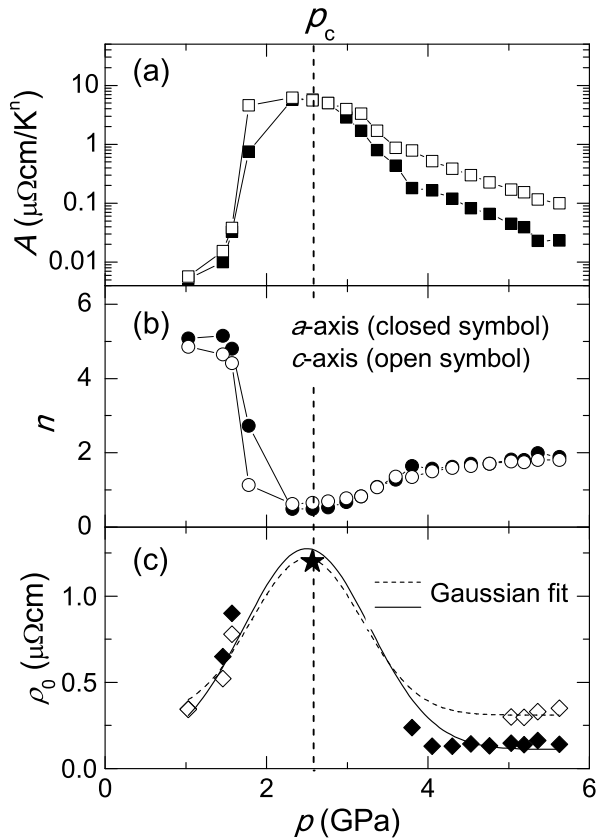}
\caption{(Color online)
(a), (b) and (c) show the pressure dependencies of the coefficient $A$, temperature exponent $n$, and residual resistivity $\rho_{\rm 0}$, respectively, obtained by fitting with the power law
$\rho(T)$ = $\rho_{\rm 0}$ + $A$$T^{n}$ to the $a$-(closed symbol) and $c$-axis (open symbol) resistivity data at low temperature.
Note that the high-field value at 2.57 GPa in panel (c) is taken from Ref. \cite{Ce115Knebel} and assumed to be isotropic.
}
\label{fig3}
\end{figure}

We now turn the attention to the low-temperature resistivity. Specifically, the $\rho_{\rm a}$ and $\rho_{\rm c}$ data are fitted by the power law $\rho$ = $\rho_{\rm 0}$ + $A$$T^{n}$ \cite{fittingrange}, where $\rho_{\rm 0}$ is the residual resistivity, $A$ the coefficient, and $n$ the temperature exponent. As shown in Fig. 3, the resulting parameters display a similar pressure dependence along different axes. At $p$ $\leqslant$ 1.57 GPa, $n$ is as large as $\sim$5, indicating dominant electron-magnon scattering due to the magnetic ordering. With increasing pressure above 1.57 GPa, since the magnetic ordering is rapidly suppressed, $n$ decreases steeply and $A$ increases accordingly. Around $p_{\rm c}$, $n$ shows a minimum of $\sim$0.6 while $A$ is enhanced by $\sim$3 orders of magnitude.
This non-Fermi liquid behavior is in good agreement with the previous results \cite{park}, pointing to the presence of quantum critical fluctuations.
Although $\rho_{\rm 0}$ obtained from the fitting is negative and hence unphysical between 2.57 and 3.80 GPa, a value near $p_{\rm c}$ can be estimated from Ref. \cite{Ce115Knebel}, in which SC can be completely suppressed by applying high magnetic fields. When plotted in Fig. 3(c), this evidences an enhanced scattering around $p_{\rm c}$, as expected \cite{rho0}.
At pressures above $\sim$4 GPa, $n$ becomes not far from the Fermi liquid value $n$ = 2.
In this pressure range, the drop in $A$ by more than one order of magnitude up to 5.63 GPa is reminiscent of the case of CeCu$_{2}$Si$_{2}$ above $p_{\rm v}$, and reflects a drastic enhancement of the 4$f$ electron interactions \cite{valencefluctuation}.

\begin{figure}
\includegraphics*[width=8.5cm]{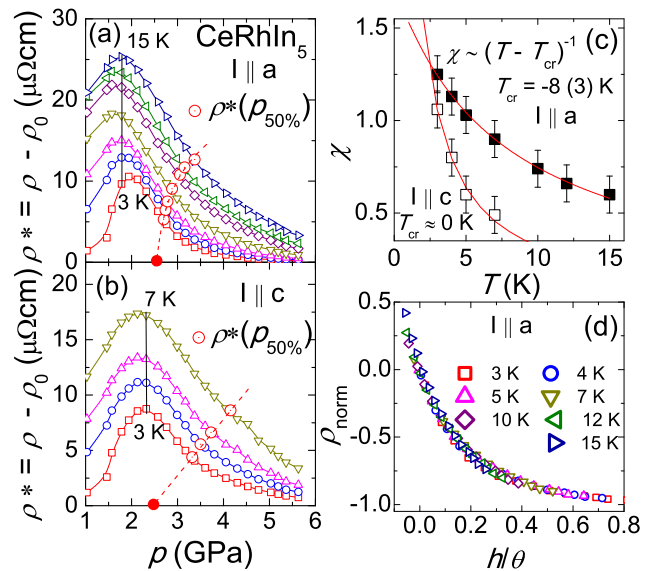}
\caption{(Color online)
(a) and (b) show the isothermal $\rho^{\rm *}$($p$) = $\rho$($p$) $-$ $\rho_{\rm 0}$($p$) for the $a$- and $c$-axis at selected temperatures, respectively.
The vertical solid lines mark the initial pressure of the valence crossover.
The solid circles denote the 50\% drop compared to the maximum $\rho^{\rm *}$ value, and
the dashed lines are extrapolations of the cricles to $p_{\rm v}$.
(c) Temperature dependencies of the slope $\chi$ for both axes (see text).
The Curie-Weiss fitting yields $T_{\rm cr}$ $\approx$ -8 K and 0 K for the $a$- and $c$-axis, respectively.
(d) Collapse of $a$-axis normalized data $\rho_{\rm norm}$ when plotted against $h$/$\theta$, where $h$ = ($p$ $-$ $p$$_{\rm 50\%}$)/$p$$_{\rm 50\%}$ and $\theta$ = ($T$ $-$ $T_{\rm cr}$)/$|$$T_{\rm cr}$$|$.
}
\label{fig4}
\end{figure}
To gain more insight, we plot the low-temperature isothermal resistivity $\rho^{\rm *}$($p$) = $\rho$($p$) $-$ $\rho_{\rm 0}$($p$) at selected temperatures in Fig. 4(a) and (b).
A maximum is observed around 1.78 and 2.32 GPa for the $a$- and $c$-axis, respectively.
At higher pressures, $\rho^{\rm *}$ decreases steeply without saturation, even in the paramagnetic state. This is also similar to that observed in CeCu$_{2}$Si$_{2}$ above 4 GPa, and, together with the rapid collapse of the $A$ coefficient shown above, provides strong evidence for the proximity to a valence CEP located at ($p_{\rm cr}$, $T_{\rm cr}$) in the $p$-$T$ plane of CeRhIn$_{5}$ \cite{seyfarth}.

\begin{figure}
\includegraphics*[width=8.5cm]{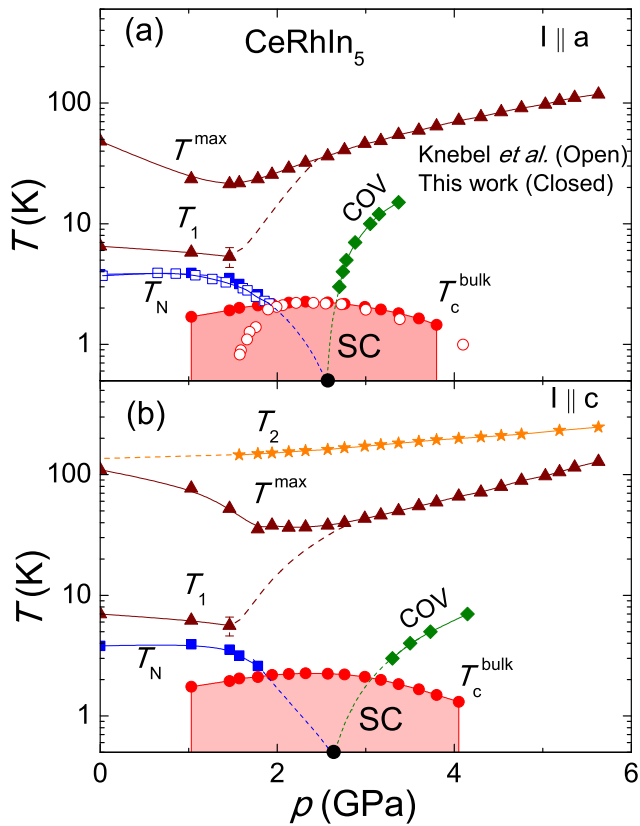}
\caption{(Color online)
Pressure-temperature phase diagram of CeRhIn$_{5}$ along (a) $a$- and (b) $c$-axis, including the characteristic temperatures  $T_{\rm 1}$, $T_{\rm 2}$ and $T^{\rm max}$.
For comparison, data from Ref.\cite{Ce115Knebel} are also include in (a).
}
\label{fig4}
\end{figure}
The existence of such a CEP can be further corroborated by a scaling analysis outlined in Ref. \cite{seyfarth}. Following the procedure, we define $p$$_{\rm 50\%}$ as the pressure corresponding to 50\% of the resistivity drop compared to the value at 1.78(2.32) GPa for the $a$($c$)-axis data, and $\rho_{\rm norm}$ as
$\rho_{\rm norm}$ = [$\rho^{\rm *}$$-$$\rho^{\rm *}$($p$$_{\rm 50\%}$)]/$\rho^{\rm *}$($p$$_{\rm 50\%}$).
As can be seen in Fig. 4(c), the slope $\chi$ = $\mid$$d$$\rho_{\rm norm}$/$dp$$\mid$ at $p$$_{\rm 50\%}$ increases with decreasing temperature, indicating that the $\rho_{\rm norm}$ decrease is getting steeper on cooling. Assuming $\chi$ $\propto$ ($T$ $-$ $T_{\rm cr}$)$^{-1}$, we obtain $T_{\rm cr}$ $\approx$ $-$8 K and 0 K for the $a$- and $c$-axis, respectively. The scaling analysis consists in plotting $\rho_{\rm norm}$ against $h$/$\theta$, where $h$ = ($p$ $-$ $p$$_{\rm 50\%}$)/$p$$_{\rm 50\%}$ and $\theta$ = ($T$ $-$ $T_{\rm cr}$)/$|$$T_{\rm cr}$$|$ are the generalized distance from the CEP. It turns out that all the $a$-axis $\rho_{\rm norm}$ isothermals below 12 K collapse on a single scaling curve \cite{caxisscaling}. This provides strong evidence for the existence of a CEP in the $p$$-$$T$ plane of CeRhIn$_{5}$. The fact that $T_{\rm cr}$ is slightly negative for the $a$-axis means that a crossover (COV) rather than a first-order transition occurs. In this respect, the temperature dependence of $p$$_{\rm 50\%}$ defines the valence COV line (see below), and its extrapolation to zero temperature yields $p_{\rm v}$($\approx$ $p_{\rm cr}$) = 2.6 $\pm$ 0.2 GPa for both cases. Notice that this $p_{\rm v}$ is determined from the results at much higher temperature than $T_{\rm N}$, yet it is nearly identical to $p_{\rm c}$.

The above results are summarized in the $p$$-$$T$ phase diagrams (PD) shown in Fig. 5. Overall, the PDs look very similar along the two crystallographic directions.
The normal-state behavior, characterized by the $T_{\rm N}$, $T_{\rm 1}$, $T^{\rm max}$ and valence COV lines, is qualitatively similar to other Ce-based Kondo lattices \cite{RenPRX}.
At low pressure, as always observed, both $T^{\rm max}$ and $T_{\rm 1}$ decreases.
For $T_{\rm 1}$, this is due to the increasing influence of the spin disorder scattering. On the other hand, the depression of $T^{\rm max}$ is ascribed to the rapid growing of the resistivity magnitude at $T_{\rm 1}$ as the Kondo temperature $T_{\rm K}$ rises. In this pressure range, $T^{\rm max}$ is primarily governed by the CF splitting $\Delta_{\rm CF}$. But at pressures above $p_{\rm v}$, since the resistivity contribution at $T_{\rm 1}$ starts to dominate, the
$T_{\rm max}$ line merges with that of $T_{\rm 1}$ and becomes an indication of $T_{\rm K}$ \cite{valencefluctuation}.

Strikingly, for both directions, the $T_{\rm N}$ and COV lines terminate at almost the same point on the horizontal-axis. In other words, the magnetic QCP at $p_{\rm c}$ nearly coincides with the valence CEP at $p_{\rm v}$, as already noted. Here we emphasize that the pressure evolution of $T_{\rm N}$ is in excellent agreement with a previous study \cite{Ce115Knebel}, although a wider superconducting window is observed in our case.
Actually, we have also performed measurements of the $a$-axis resistivity under pressure on a crystal from Thompson's group, and found identical results as those presented in this paper and notably $p_{\rm c}$ $\approx$ $p_{\rm v}$.
Hence this coincidence appears to be an intrinsic property of CeRhIn$_{5}$, and substantiates that the pressure-induced delocalization of Ce 4$f$-electron is the driving force for the collapse of the magnetic ordering in this material.

Theoretically, it has been shown that, for Ce-based periodic Anderson lattices with a large $V$, $p_{\rm c}$ and $p_{\rm v}$ are separated in the $p$$-$$T$ PD \cite{watanabe1}.
As $V$ decreases, $p_{\rm c}$ approaches $p_{\rm v}$ and finally the two pressures coincide, which is thought to correspond to the case of CeRhIn$_{5}$.
It is noted that in CeIrIn$_{5}$, which appears to have a larger $V$ than CeRhIn$_{5}$, the In nuclear quadruple resonance (NQR) measurement suggests the existence of the valence COV line near 2 GPa \cite{InNQR}, while its $p_{\rm c}$ is believed to be located at negative pressure \cite{CeIrIn5pc}.
Hence our results are in a broad agreement with previous theoretical and experimental studies.
Furthermore, according to Watanabe and Miyake \cite{watanabe1}, the coincidence of $p_{\rm c}$ and $p_{\rm v}$ consistently explains the anomalous properties observed in CeRhIn$_{5}$ by the dHvA measurement,
including the Fermi surface change and the cyclotron mass enhancement \cite{Ce115QO}.

\begin{figure}
\includegraphics*[width=8.5cm]{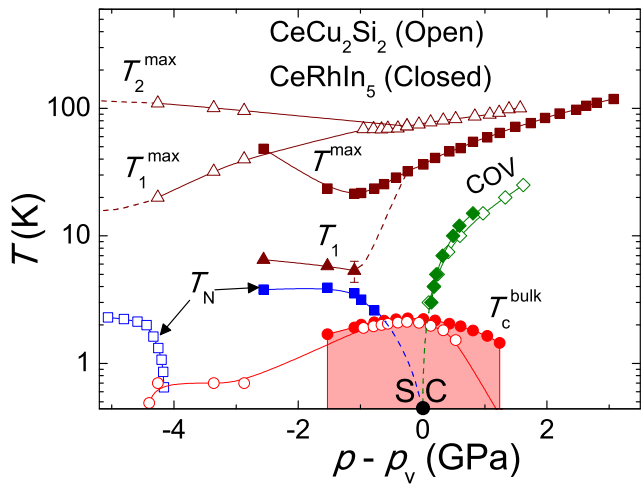}
\caption{(Color online)
Comparison of the in-plane pressure-temperature phase diagrams of CeRhIn$_{5}$ and CeCu$_{2}$Si$_{2}$ \cite{seyfarth}. Note that $p_{\rm v}$ is set as the zero on the horizontal axis for both cases.
}
\label{fig4}
\end{figure}
Another salient feature of Fig. 5 is that although $T_{\rm N}$ and $T_{c}$ are isotropic, the COV line is sharper along the $a$-axis than along the $c$-axis.
Naively, this is expected since the Ce-Ce distance is the shortest along the $a$-axis.
Hence the nucleation of valence COV develops more rapidly in this direction.
A better understanding of this issue may require further studies of the valence COV line by other probes such as NQR \cite{InNQR,CuNQR}, as well as elaborated theoretical treatments in the future.

Finally, we present in Fig. 6 a comparison between in-plane $p$$-$$T$ diagrams of CeRhIn$_{5}$ and CeCu$_{2}$Si$_{2}$.
Compared with the $T_{\rm 1}^{\rm max}$ and $T_{\rm 2}^{\rm max}$ lines of CeCu$_{2}$Si$_{2}$, the $T^{\rm max}$ and $T_{\rm 1}$ lines of CeRhIn$_{5}$ are systematically lower, which is likely due to the smaller value of the first CF splitting energy \cite{Ce115CFE}. Nevertheless, in both cases, the two lines merge in the vicinity of $p_{\rm v}$. At higher pressures, the $T_{\rm c}$ and COV lines as a function of the distance from $p_{\rm v}$ are nearly identical for these two compounds.
This is quite remarkable considering their different crystal structures, and hence points to a common superconducting pairing mechanism.
Note that, just below $p_{\rm v}$, magnetic ordering is present in CeRhIn$_{5}$, but is absent in CeCu$_{2}$Si$_{2}$.
It is thus tempting to speculate that, around the optimal pressure for superconductivity of CeRhIn$_{5}$, valence fluctuations play a more important role than spin fluctuations in the Cooper pairing, although both of them are expected to be present.

\section{IV. Conclusion}
In summary, we have studied the $a$- and $c$-axis resistivity of CeRhIn$_{5}$ under pressure up to 5.63 GPa.
A careful data analysis enables us to add the valence crossover line and to locate the CEP at 2.6 GPa and slightly negative (zero) temperature in the $p$$-$$T$ plane. For the $a$-axis, a resistivity scaling is observed, and the updated phase diagram in the COV regime is very similar to that of CeCu$_{2}$Si$_{2}$. Our results provide first experimental evidence that the magnetic QCP and valence CEP coincide with each other in CeRhIn$_{5}$, which highlights the importance of Ce-4$f$ electron delocalization in understanding the pressure evolution of magnetism and superconductivity in this material.

\section{Acknowledgement}
We acknowledge J. Flouquet for enlightening discussions, and M. Lopez and S. M\"{u}ller for technical support.
S.W. is supported by the Grant-in-Aid for Scientific Research (No. 24540378, 15K05177, and 16H10177) from the Japan Society for the Promotion of Science (JSPS), and by JASRI (No. 0046 in 2012B, 2013A, 2013B, 2014A, 2014B, and 2015A). K.M. is supported by the Grant from JSPS (No. 17K05555).


\begin{thebibliography}{99}
\expandafter\ifx\csname natexlab\endcsname\relax\def\natexlab#1{#1}\fi
\expandafter\ifx\csname bibnamefont\endcsname\relax
  \def\bibnamefont#1{#1}\fi
\expandafter\ifx\csname bibfnamefont\endcsname\relax
  \def\bibfnamefont#1{#1}\fi
\expandafter\ifx\csname citenamefont\endcsname\relax
  \def\citenamefont#1{#1}\fi
\expandafter\ifx\csname url\endcsname\relax
  \def\url#1{\texttt{#1}}\fi
\expandafter\ifx\csname urlprefix\endcsname\relax\def\urlprefix{URL }\fi
\providecommand{\bibinfo}[2]{#2}
\providecommand{\eprint}[2][]{\url{#2}}

\bibitem{Cebasedreview}
G. Knebel, D. Aoki, and J. Flouquet,
C.R. Phys. {\bf 12}, 542 (2011).

\bibitem{spinfluctuation}
N. D. Mathur, F. M. Grosche, S. R. Julian, I. R. Walker, D. M. Freye, R. K. W. Haselwimmer, and G. G. Lonzarich,
Nature (London) {\bf 394}, 39 (1998).

\bibitem{valencefluctuation}
A. T. Holmes, D. Jaccard, and K. Miyake,
Phys. Rev. B {\bf 69}, 024508 (2004).

\bibitem{seyfarth}
G. Seyfarth, A.-S. Ruetschi, K. Sengupta, A. Georges, D. Jaccard, S. Watanabe, and K. Miyake,
Phys. Rev. B {\bf 85}, 205105 (2012).

\bibitem{watanabe1}
S. Watanabe and K. Miyake,
J. Phys.: Condens. Matter {\bf 23}, 094217 (2011).

\bibitem{Miyake2}
K. Miyake and S. Watanabe,
J. Phys. Soc. Jpn. {\bf 83}, 061006 (2014).

\bibitem{Yuan}
H. Q. Yuan, F. M. Grosche, M. Deppe, C. Geibel, G. Sparn, and F. Steglich,
Science {\bf 302}, 2104 (2003).

\bibitem{RenPRX}
Z. Ren, L. V. Pourovskii, G. Giriat, G. Lapertot, A. Georges, and D. Jaccard,
Phys. Rev. X {\bf 4}, 031055 (2014).

\bibitem{Ce115discovery}
H. Hegger, C. Petrovic, E. G. Moshopoulou, M. F. Hundley, J. L. Sarrao, Z. Fisk, and J. D. Thompson,
Phys. Rev. Lett. {\bf 84}, 4986 (2000).

\bibitem{Ce115ambientSC}
G. F. Chen, K. Matsubayashi, S. Ban, K. Deguchi, and N. K. Sato,
Phys. Rev. Lett. {\bf 97}, 017005 (2006).

\bibitem{Ce115NQR}
M. Yashima, S. Kawasaki, H. Mukuda, Y. Kitaoka, H. Shishido, R. Settai, and Y. Onuki,
Phys. Rev. B {\bf 76}, 020509(R) (2007).

\bibitem{Ce115Knebel}
G. Knebel, D. Aoki, J. P. Brison, and J. Flouquet,
J. Phys. Soc. Jpn. {\bf 77}, 114704 (2008).

\bibitem{park}
T. Park, V. A. Sidorov, F. Ronning, J.-X. Zhu, Y. Tokiwa, H. Lee, E. D. Bauer, R. Movshovich, J. L. Sarrao, and J. D. Thompson,
Nature (London) {\bf 456}, 366 (2008).

\bibitem{Ce115QO}
H. Shishido, R. Settai, H. Harima, and Y. Onuki,
J. Phys. Soc. Jpn. {\bf 74}, 1103 (2005).

\bibitem{Miyake1}
S. Watanabe and K. Miyake,
J. Phys. Soc. Jpn. {\bf 79}, 033707 (2010).

\bibitem{Bridgman}
A.-S. Ruetschi and D. Jaccard,
Rev. Sci. Instrum. {\bf 78}, 123901 (2007).

\bibitem{CFE}
B. Cornut and B. Coqblin,
Phys. Rev. B {\bf 5}, 4541 (1972).

\bibitem{Ce115CFE}
A. D. Christianson, J. M. Lawrence, P. G. Pagliuso, N. O. Moreno, J. L. Sarrao,
J. D. Thompson, P. S. Riseborough, S. Kern, E. A. Goremychkin, and A. H. Lacerda,
Phys. Rev. B {\bf 66}, 193102 (2002).

\bibitem{C115MR}
A. D. Christianson, A. H. Lacerda, M. F. Hundley, P. G. Pagliuso, and J. L. Sarrao,
Phys. Rev. B {\bf 66}, 054410 (2002).

\bibitem{anistropichybridization}
Y. Matsumoto, K. Kuga, T. Tomita, Y. Karaki, and S. Nakatsuji,
Phys. Rev. B {\bf 84}, 125126 (2011).

\bibitem{fittingrange}
The temperature windows for the fitting are 1.2 K $<$ $T$ $<$ $T_{\rm N}$ at $p$ = 0, $T_{\rm c}$ $<$ $T$ $<$ $T_{\rm N}$ between 1.03 and 1.78 GPa, $T_{\rm c}$ $<$ $T$ $<$ 6 K between 1.94 and 3.80 GPa, and 1.2 K $<$ $T$ $<$ 6 K above 4.02 GPa. In addition,

\bibitem{rho0}
K. Miyake and O. Narikiyo,
J. Phys. Soc. Jpn. {\bf 71}, 867 (2002).

\bibitem{caxisscaling}
Note that, for the $c$-axis case, $\theta$ tends to $\infty$ since $T_{\rm cr}$ $\sim$ 0.
Hence $h$/$\theta$ is not physically meaningful.

\bibitem{InNQR}
M. Yashima, N. Tagami, S. Taniguchi, T. Unemori, K. Uematsu, H. Mukuda, Y. Kitaoka, Y. \={O}ta, F. Honda, R. Settai, and Y. \={O}nuki,
Phys. Rev. Lett. {\bf 109}, 117001 (2012).


\bibitem{CeIrIn5pc}
P. G. Pagliuso, R. Movshovich, A. D. Bianchi, M. Nicklas, N. O. Moreno, J. D. Thompson, M. F. Hundley, J. L. Sarrao, and Z. Fisk,
Physica B {\bf 312-313}, 129 (2002).


\bibitem{CuNQR}
T. C. Kobayashi, K. Fujiwara, K. Takeda, H. Harima, Y. Ikeda, T. Adachi, Y. Ohishi, C. Geibel, and F. Steglich,
J. Phys. Soc. Jpn. {\bf 82}, 114701 (2013).

\end{thebibliography}
\end{document}